\documentclass[twocolumn,floats,prd,amssymb,floatfix,nofootinbib,balancelastpage,superscriptaddress,amsmath]{revtex4}

\pdfoutput=1

\usepackage{epsfig}
\usepackage{latexsym}
\usepackage{hyperref}
\usepackage{graphics,graphicx,color,amssymb,float,amsmath}
\usepackage{pstricks}
\usepackage{color,ulem}
\usepackage{mathrsfs}
\usepackage{xspace}

\def\beqn{\begin{eqnarray}} 
\def\eeqn{\end{eqnarray}} 
\def\be{\begin{equation}}
\def\ee{\end{equation}}
\def\nn{\nonumber}

\newcommand{\tax}{{\tilde{a}}}

\newcommand{\newc}{\newcommand}
\newc{\half}{\frac{1}{2}}
\newc{\Lam}{{\bf \Lambda}}
\newc{\ltau}{\lambda_\tau}
\newc{\lt}{\lambda_t}
\newc{\lb}{\lambda_b}
\newc{\kap}{{\bf \kappa}}
\newc{\lae}{{\Lam}_E}
\newc{\lad}{{\Lam}_D}
\newc{\lau}{{\Lam}_U}
\newc{\lame}[1]{{\Lam}_{E^{#1}}}
\newc{\lamhe}[1]{{\h}_{E^{#1}}}
\newc{\lamhed}[1]{{\h}_{E^{#1}}^\dagger}
\newc{\lamhd}[1]{{\h}_{D^{#1}}}
\newc{\lamhdd}[1]{{\h}_{D^{#1}}^\dagger}
\newc{\lamhu}[1]{{\h}_{U^{#1}}}
\newc{\lamhud}[1]{{\h}_{U^{#1}}^\dagger}
\newc{\lamd}[1]{{\Lam}_{D^{#1}}}
\newc{\lamu}[1]{{\Lam}_{U^{#1}}}
\newc{\lamet}[1]{{\Lam}_{E^{#1}}^T}
\newc{\lamdt}[1]{{\Lam}_{D^{#1}}^T}
\newc{\lamut}[1]{{\Lam}_{U^{#1}}^T}
\newc{\lames}[1]{{\Lam}_{E^{#1}}^*}
\newc{\lamds}[1]{{\Lam}_{D^{#1}}^*}
\newc{\lamus}[1]{{\Lam}_{U^{#1}}^*}
\newc{\lamed}[1]{{\Lam}_{E^{#1}}^\dagg}
\newc{\lamdd}[1]{{\Lam}_{D^{#1}}^\dagg}
\newc{\lamud}[1]{{\Lam}_{U^{#1}}^\dagg}
\newc{\lam}{{\bf \lambda}}
\newc{\lamp}{{\bf \lambda}^{\prime}}
\newc{\lampp}{{\bf \lambda}^{\prime\prime}}
\newc{\Y}{{\bf Y}}
\newc{\h}{{\bf h}}
\newc{\meee}{{{\rm {\bf  m}}_e}}
\newc{\mdee}{{{\rm {\bf  m}}_d}}
\newc{\myew}{{{\rm {\bf m}}_u}}
\newc{\ye}{{\Y}_E}
\newc{\he}{{\h}_E}
\newc{\hed}{{\h}_E^\dagger}
\newc{\yd}{{\Y}_D}
\newc{\hd}{{\h}_D}
\newc{\hdd}{{\h}_D^\dagger}
\newc{\yu}{{\Y}_U}
\newc{\hu}{{\h}_U}
\newc{\hud}{{\h}_U^\dagger}
\newc{\yes}{{\Y}_E^*}
\newc{\yds}{{\Y}_D^*}
\newc{\yus}{{\Y}_U^*}
\newc{\yet}{{\Y}_E^T}
\newc{\ydt}{{\Y}_D^T}
\newc{\yut}{{\Y}_U^T}
\newc{\yed}{{\Y}_E^\dagg}
\newc{\ydd}{{\Y}_D^\dagg}
\newc{\yud}{{\Y}_U^\dagg}
\newc{\dagg}{\dagger}

\newc{\fPQ}{f_{\rm{PQ}}}
\newc{\gsim}{\stackrel{>}{\sim}}
\newc{\lsim}{\stackrel{<}{\sim}}

\newc{\htext}[1]{{\color{red}  #1}}
\newc{\bltext}[1]{{\color{blue}  #1}}

\newc{\kpvertex}{\Phi_{\tax}H_2H_1}
\newc{\kppvertex}{\Phi_{\tax}L_iH_2}

\begin{document}
 \DeclareGraphicsExtensions{.pdf,.png,.gif,.jpg}

\title{Heavy concerns about the light axino explanation of the 3.5 keV X-ray line}

\author{ Stefano Colucci}
\email{colucci@th.physik.uni--bonn.de}
\affiliation{Physikalisches Institut der Universit\"at Bonn, Bethe Center
for Theoretical Physics, \\ Nu{\ss}allee 12, 53115 Bonn, Germany}

\author{Herbi K. Dreiner}
\email{dreiner@th.physik.uni--bonn.de}
\affiliation{Physikalisches Institut der Universit\"at Bonn, Bethe Center
for Theoretical Physics, \\ Nu{\ss}allee 12, 53115 Bonn, Germany}

\author{Florian Staub}
\email{florian.staub@cern.ch}
\affiliation{Theory Division, CERN, 1211 Geneva 23, Switzerland}

\author{Lorenzo Ubaldi}
\email{ubaldi.physics@gmail.com}
\affiliation{Raymond and Beverly Sackler School of Physics and Astronomy, \\
 Tel-Aviv University, Tel-Aviv 69978, Israel}

 \begin{picture}(0,0)
  \put(0,12){CERN-PH-TH-2015-163} \\
  \put(0,0){BONN-TH-2015-10}
 \end{picture}

\begin{abstract}
An unidentified 3.5 keV line from X-ray observations of galaxy clusters has been reported recently. Although still under scrutiny, decaying dark matter could be responsible for this signal. We investigate whether an axino with a mass of 7 keV could explain the line, keeping the discussion as model independent as possible. We point out several obstacles, which were overlooked in the literature, and which make the axino an unlikely candidate. The only viable scenario predicts a light metastable neutralino, with a mass between 0.1 and 10 GeV and a lifetime between $10^{-3}$ and $10^4$ s.

\end{abstract}

\maketitle

\section{Introduction}
\label{sec:Introduction}

There has been interest recently in an unidentified 3.5 keV line in X-ray observations of galaxy 
clusters~\cite{Boyarsky:2014jta, Bulbul:2014sua}. Despite its physical origin being subject to debate~\cite{Urban:2014yda, Jeltema:2014qfa, Bulbul:2014ala, Boyarsky:2014paa, Jeltema:2014mla,Carlson:2014lla}, there is still some room to speculate 
that decaying dark matter is responsible for this signal. The first obvious dark matter (DM) candidate in this context is 
a 7 keV sterile neutrino~\cite{Boyarsky:2014jta}, but many other alternatives have been proposed. Some authors have pointed 
out that a decaying axino could explain the line~\cite{Liew:2014gia, Kong:2014gea, Choi:2014tva, Dutta:2014saa}. In this paper we 
examine carefully the conditions under which a 7 keV axino would produce the observed X-ray line. Due to the large number of 
parameters at disposal in supersymmetric models, it is hard to exclude with certainty the axino scenario. However we show that 
various constraints leave almost no room available in the parameter space of these models. Therefore we deem the axino an unlikely candidate to explain the line. The authors of Ref.~\cite{Kolda:2014ppa} also mentioned the axino as an unnatural explanation of the X-ray line. In this work we elaborate on the physical arguments that lead to such a conclusion.

There are a few reasons why it is appealing to consider a model with a light axino. First, introducing the axion multiplet 
(axion, axino, saxion) in models of supersymmetry (SUSY) solves the strong CP problem~\cite{Peccei:2006as}. Second, 
the axion and the axino can both be DM candidates. Third, if the axino has a mass in the keV range, it is warm DM 
and it could help reconcile some small-scale structure issues~\cite{VillaescusaNavarro:2010qy, Viel:2013fqw, Schneider:2011yu} 
of cold DM. In the context of R-parity violating (RPV) SUSY \cite{Dreiner:1997uz}, a light axino is unstable. Its lifetime is still longer than the age of the universe, but it can decay into a neutrino and a photon. 
It is this channel that would produce the 3.5~keV line. RPV models also have the virtue of explaining the null-results for 
SUSY searches at the LHC without introducing a little hierarchy problem: the limits on the masses of the sparticles 
become much weaker~\cite{Allanach:2012vj,Asano:2012gj,Chamoun:2014eda,Franceschini:2012za,Evans:2012bf}. The axino could also produce an X-ray 
photon in R-parity conserving SUSY. This is possible if the lightest neutralino is very light or even massless. The axino can then decay into the neutralino plus a photon~\cite{Dutta:2014saa}. We comment 
also on this possibility, below.

This letter is organized as follows. In sec.~\ref{sec:relic}, we review how the axino abundance depends on the reheating 
temperature, in sec.~\ref{sec:xray} we discuss various constraints that the 3.5 keV line puts on models with a decaying 
axino. We conclude in sec.~\ref{sec:conclusion}.

\section{Axino relic density}
\label{sec:relic}
Any supersymmetric model with an axino contains also an axion. The latter, as an invisible axion,  
is always a good DM candidate, while the former is suitable only if it is sufficiently long lived. We are here interested in a 
scenario where the axino constitutes almost the entire DM budget of the universe. This is typically realized for low values 
of the axion decay constant ($f_a \sim 10^{10}\,$GeV), in which case the axion DM component is suppressed and can 
be neglected. See {\it e.g.} Ref.~\cite{Kawasaki:2013ae} for a review on axion DM.

The axino as a DM candidate has been widely explored in the literature~\cite{Rajagopal:1990yx,Covi:2001nw, Covi:2009pq, Strumia:2010aa, Choi:2011yf,Choi:2013lwa, Bae:2011jb, Bae:2011iw}. 
There are two axion/axino scenarios of interest: the Dine-Fischler-Srednicki-Zhitnitsky (DFSZ) 
model~\cite{Dine:1981rt, Zhitnitsky:1980tq}, and the  Kim-Shifman-Vainshtein-Zakharov 
(KSVZ) model~\cite{Kim:1979if, Shifman:1979if}. As far as the axino is concerned, the important difference between the two lies 
in the mass, $M_\Phi$, of the heaviest Peccei-Quinn (PQ) charged and gauge-charged matter supermultiplet in the 
model~\cite{Bae:2011jb}. In the DFSZ case $M_\Phi$ is the Higgsino mass, of order $\lesssim\,$TeV, while in the KSVZ 
case it is the mass of the heavy vectorlike quarks, typically of order $f_a\gtrsim10^9\,$GeV. This difference leads to 
two different relic axino abundances as well as to a different dependence on the reheating temperature, 
$T_{\rm RH}$. For DFSZ models we have~\cite{Choi:2011yf}
\be 
\Omega^{\rm DFSZ}_{\tilde a} h^2 \simeq 0.78  \left( \frac{m_{\tilde a}}{7 \ {\rm keV}} \right) \left( \frac{10^{10} \ {\rm GeV}}{f_a} 
\right)^2 \, ,
\label{eq:axinoabundance}
\ee
which does not depend on $T_{\rm RH}$, as long as $T_{\rm RH}$ is larger than\footnote{The exact value of $T_{\rm RH}$ 
 depends on the SUSY spectrum, but it is expected to lie in the TeV range.}\! roughly 1~TeV 
 \cite{Bae:2011iw}.  The relic density drops very quickly for lower values of $T_{\rm RH}$ (see {\it e.g.} Fig.~3 
 in Ref.~\cite{Choi:2011yf}). From eq.~\eqref{eq:axinoabundance} we see that for a 7 keV DFSZ axino 
 with $f_a < 10^{10}$ GeV the reheating temperature has to be below 1~TeV in order to avoid 
 overabundance. Recall, obervationally $\Omega_{\mathrm{DM}} h^2=0.1199\pm0.0027$ \cite{Ade:2013zuv}.

For the KSVZ case~\cite{Choi:2011yf}:
\begin{align}
\Omega_{\tilde a}^{\rm KSVZ} h^2 = 6.9 \times 10^{-3} & \left(  \frac{m_{\tilde a}}{7 \ {\rm keV}} \right) \left( \frac{10^{10} \ 
{\rm GeV}}{f_a} \right)^2 \nn \\
& \times \left(\frac{T_{\rm RH}}{10^3 \ {\rm GeV}} \right) \, , \label{eq:KSVZab}
\end{align}
which is valid for  $1 \ {\rm TeV}<T_{\rm RH} < M_\Phi \sim f_a$. Also in this case the abundance drops very quickly for reheating 
temperatures below about 1~TeV.

\subsection{Including a light gravitino}
Without tuning and/or cancellations, the axino is expected to be heavier than the 
gravitino~\cite{Cheung:2011mg,Dreiner:2014eda}. This puts further constraints on our scenario. First, the axino can 
decay into a gravitino and an axion with a lifetime \cite{Cheung:2011mg}
\be
\tau_{\tilde a \to \tilde G + a} \simeq 6\cdot10^{28} \ {\rm s} \left( \frac{m_{3/2}}{{\rm keV}} \right)^2 \left( \frac{\rm 7 \,keV}
{m_{\tilde a}} \right)^5 \, .
\ee
Here $m_{3/2}$ denotes the gravitino mass.
If the 7 keV axino is to explain the X-ray line we must require its lifetime to be roughly greater than the age of the universe, $\tau_{\tilde a \to \tilde G + a} > 10^{18}\,$s, otherwise it would 
have decayed away. This implies $m_{3/2} > 10^{-5}\,$keV.  A gravitino lighter than roughly 
100 eV is a hot dark matter candidate. Its relic abundance~\cite{Kolb:1990vq},
\be \label{eq:Ghot}
\Omega_{\tilde G}^{\rm HDM} h^2 \sim 0.1\,\frac{m_{3/2}}{100 \ {\rm eV}} \, , \quad {\rm for} \ m_{3/2} < 100 \ {\rm eV} \, ,
\ee
is constrained to be less than 3\% of the total DM abundance~\cite{Viel:2005qj}, which implies $m_{3/2} \lesssim1\,$eV. 

In the range 1~keV $< m_{3/2} < 7$ keV, the gravitino is a warm dark matter candidate and its abundance 
depends on $T_{\rm RH}$~\cite{Bolz:2000fu}: 
\be \label{eq:Gabundance}
\Omega_{\tilde G} h^2 = 0.27 \left( \frac{T_{\rm RH}}{100 \ {\rm GeV}} \right) \left( \frac{\rm keV}{m_{3/2}} \right), \,{\rm for} \ 
m_{3/2} > 1\, {\rm keV} \, .
\ee
In this case we require a reheating temperature below $1-10$ GeV so that the gravitino contribution to the DM density
is much smaller than that of the axino. However for such low values of $T_{\rm RH}$ the axino relic density, 
even with $f_a \sim 10^9$ GeV, is highly suppressed, $\Omega_{\tilde a} h^2 \ll 0.12$. Therefore, we exclude this case.

Thus if the axino constitutes most of the DM and produces the 3.5 keV X-ray line, the gravitino mass is 
restricted to the small window
\be \label{eq:Grange}
10^{-2} \ {\rm eV} < m_{3/2} < 1 \ {\rm eV} \, .
\ee

As both the axino and the gravitino are in the keV range and below, one might worry about their contribution to the 
relativistic degrees of freedom at Big Bang Nucleosynthesis (BBN). This turns out not to be a problem in most cases. When the universe reheats 
to 10 - 100 GeV, both the axino and the gravitino are out of thermal equilibrium~\cite{Choi:2011yf}. The subsequent 
annihilations of SM particles heat up the photon bath so that the photon temperature at BBN is higher than the respective temperatures of the axino and the gravitino. As a consequence their 
contribution to $N_{\rm eff}$ is well within the bound. If there is also a very light or massless neutralino in the spectrum 
one has to worry about the constraint from $N_{\rm eff}$, as we discuss later.

\section{The axino and the X-ray line}
\label{sec:xray}

To explain the 3.5 keV X-ray line via the decay of a DM particle $X$, one needs a decay rate of $\Gamma_
{X \to \gamma + \dots} \sim (10^{28} \ {\rm s})^{-1} \sim 10^{-53}\,$GeV \cite{Boyarsky:2014jta}, assuming that the 
decaying DM constitutes 100\% to the relic abundance. If the decaying DM is a fraction 
$k <1$ of the total DM, then the corresponding decay rate has to increase by $1/k$ to explain the signal. 

There are two scenarios for a 7~keV axino to produce the 3.5 keV X-ray line, where R-parity is respectively 
conserved or violated. In either case the starting point is the following Lagrangian for the coupling of the axino to 
the SM gauge bosons and their gaugino supersymmetric partners:
\begin{align}
\mathcal{L}_{a\lambda V} = & \frac{\tilde a}{16 \pi^2 f_a}  \sigma_{\mu\nu} \Big( g_1^2 C_{aBB} \tilde B  B^{\mu\nu} 
+ \nonumber \\  
& \quad g_2^2 C_{aWW} \tilde W^a   W^{a \mu\nu} + g_3^2  \tilde G^\alpha   G^{\alpha \mu\nu} \Big) \, .
\end{align}
Here $\tilde a$ is the axino mass eigenstate, while the gauginos and gauge fields are gauge eigenstates; $g_i$ are 
the gauge couplings. These interactions are the supersymmetric version of those of the axion with gauge fields in the 
low energy (non-supersymmetric) Lagrangian~\cite{Georgi:1986df}. The coefficients $C$ are model dependent 
and are typically of order one~\cite{Kim:2008hd}.

\subsection{R-parity conserving SUSY}
\label{sec:RPC}
In the R-parity conserved case, if the bino is lighter than the axino, the X-ray line could be produced by the decay 
$\tilde a \to \tilde B + \gamma$. This was pointed out in Ref.~\cite{Dutta:2014saa}. The axino partial decay rate 
with a massless bino is
\begin{align} \label{eq:dutta}
\Gamma_{\tilde a \to \tilde B + \gamma} & =  \frac{1}{128 \pi^3} \frac{m_{\tilde a}^3}{f_a^2} C_{aBB}^2 \left( \frac{g_1^2}{4 \pi} \right)^2 \cos^2 \theta_W \\
& \sim 7\times 10^{-52} \ {\rm GeV} \left( \frac{m_{\tilde a}}{7 \ {\rm keV}} \right)^3 \left( \frac{10^{14} \ {\rm GeV}}{f_a} \right)^2 \, .
\end{align}
Note, a massless bino is consistent with all data provided the sfermions are heavy enough 
\cite{Dreiner:2003wh,Dreiner:2009ic,Dreiner:2009er}. To match the rate needed for the X-ray line this scenario 
requires $f_a > 10^{14}$ GeV. This immediately excludes the DFSZ axino, whose relic abundance would be 
much too low. 

From eq.~\eqref{eq:KSVZab}, a KSVZ axino with $T_{\rm RH} \sim 10^{12}$ GeV would seem viable. However this 
scenario is strongly disfavored by two arguments. First, the abundance of axions produced via the misalignment 
mechanism would be far too high~\cite{Kawasaki:2013ae} for $f_a \sim 10^{14}$ GeV, unless one tunes the initial 
misalignment angle to very 
small values. Second, a massless bino together with a light gravitino would contribute to the relativistic degrees of 
freedom~\cite{Dreiner:2011fp}, giving a value $\Delta N_{\rm eff} > 1$, in strong tension with BBN and with the 
data from the Cosmic Microwave Background (CMB)~\cite{Planck:2015xua}.

\subsection{R-parity violating SUSY}
In the context of R-parity violation, the terms $\epsilon_i L_i H_u,\;i=e,\mu,\tau$ in the superpotential 
introduce mixing among the neutrinos and the Higgsinos. The modified scalar potential also results in 
non-zero sneutrino vacuum expectation values (VEVs), which introduce mixing between the neutrinos and the bino 
and the neutral wino, respectively. In this case the
RPC axino decay in sec.~\ref{sec:RPC} automatically includes the decay channel $\tilde a \to \nu_i + 
\gamma$. The new partial decay rate is simply modified by the appropriate mixing
angles\footnote{Note this decay would also occur with pure trilinear R-parity violation via the resulting sneutrino vev's
\cite{Allanach:2003eb}.}
 \begin{align}
\Gamma_{\tilde a \to \nu_i + \gamma} & =  \frac{1}{128 \pi^3} \frac{m_{\tilde a}^3}{f_a^2} \left[r_{\nu_i \tilde B}^2 C_{aBB}^2 
\left( \frac{g_1^2}{4 \pi} \right)^2 \cos^2 \theta_W \right.
\nonumber \\ 
& \qquad \left. + r_{\nu_i \tilde W}^2 C_{aWW}^2 \left( \frac{g_2^2}{4 \pi} \right)^2 \sin^2 \theta_W \right] \\
& \sim   7\times 10^{-42} \ {\rm GeV} \  (r_{\nu_i \tilde B}^2 + 3 r_{\nu_i \widetilde W}^2) \nn \\
& \qquad \times \left( \frac{m_{\tilde a}}{7 \ {\rm keV}} \right)^3 \left( \frac{10^{9} \ {\rm GeV}}{f_a} \right)^2 \, . \label{eq:GRPV}
\end{align}
Here $r_{\nu_i \tilde B}$ ($r_{\nu_i \widetilde W}$) parametrizes the mixing of the neutrino mass eigenstate with 
the gaugino gauge eigenstate $\tilde B$ ($\widetilde W^0$). The lifetime of the axino to explain the X-ray 
line requires $r^2_{\nu_i \tilde B}$ ($r^2_{\nu_i \tilde W}$) to be of order 
$10^{-12}$ for $f_a$ fixed at its lowest possible value~\cite{Raffelt:2006cw}, $10^9\,$GeV. 

One the outstanding features of RPV models with lepton number violation, is that they automatically provide for 
massive neutrinos. Assuming that the neutrino masses solely  arise from the RPV sector, we can estimate 
bounds on the mixings $r_{\nu_i (\tilde B,\widetilde W)}$ as follows.  Neglecting loop contributions, the terms 
$\epsilon_i L_i H_u$ lead to one massive neutrino 
\cite{Hirsch:1998kc,Banks:1995by,Hirsch:2004he} and two non-vanishing lepton mixing angles, which we take 
to be $\theta_{13}$ and $\theta_{23}$ \cite{Hirsch:1998kc}. The neutrino mass
is given in terms of the model parameters as
\be 
\label{eq:nu3mass}
m_{\nu_3} = \frac{g_2^2 M_1 + g_1^2 M_2}{4 \det \mathcal{M}_{\chi_0}} |\vec{\Lambda}|^2 \, ,
\ee
where \cite{Hirsch:1998kc}
\beqn
\Lambda_i & \equiv & \mu \omega_i - v_d \epsilon_i \, ,\quad i=e,\mu,\tau
\eeqn
are the alignment parameters and  $\omega_i \equiv \langle \tilde \nu_i \rangle$ are the sneutrino VEVs.
Furthermore $\det \mathcal{M}_{\chi_0}$ denotes the determinant of the $4\times 4$ neutralino 
sub-mass-matrix of the MSSM
\begin{equation}
\det \mathcal{M}_{\chi_0} \equiv  -\mu^2 M_1 M_2  
+ \frac{1}{2} \mu v_u v_d (g_2^2 M_1  + g_1^2 M_2) \, ,
\end{equation}
where $v_u \equiv \langle H_u \rangle$, $v_d \equiv \langle H_d \rangle$ are the Higgs VEVs. The 
parameters $\Lambda_i$ are related to the two remaining neutrino mixing angles~\cite{Hirsch:1998kc}:
\be~\label{eq:thetas}
\tan \theta_{13} = - \frac{\Lambda_e}{(\Lambda_\mu^2 + \Lambda_\tau^2)^{1/2} } \, , \qquad \tan \theta_{23} = 
\frac{\Lambda_\mu}{\Lambda_\tau} \, .
\ee
These angles are measured~\cite{Agashe:2014kda} to be $\theta_{13} \sim \pi / 20$ and $\theta_{23} \sim \pi / 4$, 
thus we have $\Lambda_\mu = \Lambda_\tau \equiv \Lambda$ and $\Lambda_e = 0.23\, \Lambda$. The 
cosmological bound on the sum of the neutrino masses~\cite{Planck:2015xua}, $\sum_i m_{\nu_i} < 0.23$ eV, in 
our case amounts to a bound on the single massive neutrino, and thus
\be
\Lambda < (3.2 \times 10^{-13} \ {\rm TeV})^{1/2} \left( \frac{\det \mathcal{M}_{\chi_0}}{g_2^2 M_1  + g_1^2 M_2}  
\right)^{1/2} \!\!.
\ee
From Refs.~\cite{Hirsch:1998kc, Hirsch:2004he} we can work out the mixings analytically, in terms of the
supersymmetric parameters. For instance
\be
r_{\nu_2 \tilde B}  =  \frac{g_1 M_2 \Lambda_\mu \Lambda_\tau \mu}{\sqrt{ \Lambda_\mu^2 + \Lambda_\tau^2}\cdot\det \mathcal{M}_{\chi_0} 
} 
\simeq   \frac{g_1 M_2  \Lambda \mu}{\sqrt{2}\det \mathcal{M}_{\chi_0} 
} \, ,
\ee
and similar expressions for the other mixings. Then the bound on $\Lambda$ translates into
\beqn
r_{\nu_1 \tilde B}^2 & < & (1.1\times 10^{-14} \ {\rm TeV}) \mathcal{M}^{-1} \\
r_{\nu_2 \tilde B}^2 & < & (4.6\times 10^{-13} \ {\rm TeV}) \mathcal{M}^{-1} \\
r_{\nu_3 \tilde B}^2 & < & (2.8\times 10^{-16} \ {\rm TeV})  \mathcal{M}^{-1} \, ,
\eeqn
with 
\be
\mathcal{M}^{-1} \, \equiv \,  \frac{g_1^2 M_2^2 \mu^2}{(g_2^2 M_1 + g_1^2 M_2)\det \mathcal{M}_{\chi_0}
} \, ,
\ee
and similar bounds for $r_{\nu_1 \widetilde W}^2$. Given the null SUSY searches at the LHC so far, it is reasonable to expect 
the parameters $M_1, M_2, \mu$ to be of order TeV or larger, and $\det \mathcal{M}_{\chi_0} \sim {\rm TeV}^4$, in absence 
of cancellations. In this case the bounds simplify to
\beqn
r_{\nu_1 \tilde B}^2 & < & 2.3 \times 10^{-15} \, ,\\
r_{\nu_2 \tilde B}^2 & < & 9.2 \times 10^{-14} \, ,\\
r_{\nu_3 \tilde B}^2 & < & 5.6 \times 10^{-17} \, .
\eeqn
These mixings are very small and thus the resulting axino decay rate too slow to explain 
the X-ray line.

Perhaps the assumption that all the SUSY parameters and masses are at least around a TeV is too strict. Suppose 
that the lightest neutralino, $\chi^0_1$, has a mass of order GeV and the other neutralinos are at the TeV scale. 
Then $\det \mathcal{M}_{\chi_0} \sim 10^{-3} \ {\rm TeV}^4$ and the bound on $r_{\nu_2 \tilde B}^2$ becomes of order 
$10^{-10}$, which is enough to fit the line. However this scenario faces another problem. A GeV neutralino has two decay channels: one into an axino and a photon with
\begin{align}\label{eq:GammaChi}
\Gamma_{\chi^0_1 \to \tilde a + \gamma} & \simeq  \frac{1}{128 \pi^3} \frac{m_{\tilde B}^3}{f_a^2} C_{aBB}^2 \left( \frac{g_1^2}{4 \pi} \right)^2 \cos^2 \theta_W  \, , \nn \\
\tau_{\chi^0_1 \to \tilde a + \gamma} & \sim 60 \ {\rm s} \left( \frac{ {\rm GeV}}{m_{\chi^0_1}} \right)^3 \left( \frac{f_a}{10^{9} \ {\rm GeV}} \right)^2 \, ;
\end{align}
the other into a neutrino plus leptons (or 
pions) via an off-shell $Z$ boson, with~\cite{Dreiner:1991pe}
\begin{align}
 \Gamma_{\chi^0_1 \to \nu l^+ l^-} & \simeq \frac{r_{\nu_i \tilde B}^2 \alpha^2}{1024 \pi^3} \frac{m_{\chi^0_1}^5}{M_Z^4} \, , \nn \\
 \tau_{\chi^0_1 \to \nu l^+ l^-} & \sim 60 \ \text{s} \ \left(\frac{10^{-10}}{r_{\nu_i \tilde B}^2} \right) \left(\frac{\rm GeV}{m_{\chi^0_1}} \right)^5 \, .
\end{align}
For $m_{\chi^0_1} < 1$ GeV the dominant decay mode is into axino plus photon, while for $m_{\chi^0_1}>1$ GeV it is into neutrino plus leptons.  
These decay rates are low enough that the neutralino freezes out before decaying. As a weakly interacting particle 
(WIMP), at freeze-out it has roughly the right DM relic abundance, $\Omega_{\chi^0_1} h^2 \sim 0.1$. If $\chi^0_1$ 
is mostly bino, it will be slightly overabundant, while if it is mostly wino or higgsino it will be 
underabundant~\cite{Martin:1997ns}. In either case our subsequent conclusions do not change. 
When $\chi^0_1$ decays it produces energetic photons, which are subject to constraints from BBN and CMB.
The window $10 {\rm~MeV} < m_{\chi^0_1} < 100 {\rm~MeV}$, corresponding to $3\times 10^4 {\rm~s} < \tau_{\chi^0_1} < 3\times 10^7 {\rm~s}$, is excluded by bounds from photodestruction of $D$ and photoproduction of $D+^3He$~\cite{Ellis:1990nb}. The window $300 {\rm~keV} < m_{\chi^0_1} < 10 {\rm~MeV}$, corresponding to $3\times 10^7 {\rm~s} < \tau_{\chi^0_1} <  10^{12} {\rm~s}$, is excluded by bounds from CMB spectrum distortions~\cite{Hu:1992dc, Ellis:1990nb}. The window $7 {\rm~keV} < m_{\chi^0_1} < 300 {\rm~keV}$, with a lifetime $\tau_{\chi^0_1} > 10^{12} {\rm~s}$, is excluded by CMB constraints on late decaying particles~\cite{Slatyer:2012yq}.
For $m_{\chi^0_1} < 7 {\rm~keV}$ the neutralino is lighter than the axino
and we are back to the situation of the R-parity conserving scenario, which is strongly disfavored as we discussed in 
the previous section. 

The window $100 {\rm~MeV} < m_{\chi^0_1} < 10 {\rm~GeV}$, corresponding to $10^{-3} {\rm~s} < \tau_{\chi^0_1} < 3\times 10^4 {\rm~s}$, cannot be as easily excluded, so it is viable in principle. In this range most of the neutralinos decay around BBN time, which could be in some tension with the success of BBN itself. However a detailed study, beyond the scope of the current work, is needed to come to a definite conclusion.  
For $m_{\chi_1^0}>10$ GeV one quickly hits the bound from neutrino masses. 

We conclude that also the RPV scenario is strongly disfavored.

\section{Summary}
\label{sec:conclusion}
Motivated by recent claims of detection of an X-ray line at 3.5 keV we have investigated whether a decaying axino with a 7 keV mass could explain the signal. The R-parity conserving scenario is strongly disfavored mostly because it requires a very light bino, which together with a light gravitino would contribute to the relativistic degrees of freedom, $\Delta N_{\rm eff}>1$, in contradiction with BBN and CMB bounds. RPV models require a mixing neutrino-bino or neutrino-wino which is typically too large and excluded by the cosmological bound on neutrino masses. To evade this bound one is forced to take the lightest neutralino with a mass around a GeV or below. Such a neutralino is long-lived and decays into energetic photons. If it is lighter than 100 MeV the scenario is excluded by BBN and CMB constraints. These arguments are generic, model independent, and make the 7 keV axino a very unlikely candidate for the observed 3.5 keV line.

If the lightest neutralino has a mass between 100 MeV and 10 GeV, the RPV scenario is still viable. 
The corresponding lifetime of the neutralino ranges from $\tau_{\chi^0_1}=\mathcal{O}(10^{-3}\,$s), for $m_{\chi^0_1} = 10$ GeV, 
to $\tau_{\chi^0_1}=\mathcal{O}(10^4\,$s), for
$m_{\chi^0_1} =100$ MeV. This is effectively stable for
collider physics. Such a neutralino is very hard to observe in the laboratory, with no present bounds~\cite{Dreiner:2009ic}.
 The low mass range is excluded by supernova cooling if the selectron is lighter
than about 500 GeV~\cite{Dreiner:2003wh}. The proposed SHiP facility at CERN is most likely not 
sensitive to these lifetimes, due to the restricted geometry~\cite{Alekhin:2015byh}. A direct measurement
at an $e^+e^-$ linear collider is also unlikely~\cite{Conley:2010jk}. 
This range of lifetimes is too short for possible astrophysical signatures. Therefore it is hard to verify whether this particular scenario
is realized or excluded.

It might still be possible, at the expense of fine tuning, to find other corners of parameter space where the axino has the right abundance and decay rate to fit the line. One would have to do a numerical scan for a specific model, which is beyond the scope of this work. If such a corner were found, our arguments suggest that it would have a low axion decay constant, $f_a \sim 10^9$ GeV, at the edge of the astrophysical bound, and a low reheating temperature, below a TeV (for a KSVZ model) or even around a few tens of GeV (for a DFSZ model). This is troublesome for most of the proposed baryogenesis mechanisms. 

\subsection*{Note added}
As this work was being completed, Ref.~\cite{2015arXiv150704619P} appeared, which casts further doubts on the DM interpretation of the 3.5 keV line.

\acknowledgments
HD and SC acknowledge the DFG SFB TR
33 `The Dark Universe' for support throughout this work.
LU is supported by the I-CORE Program of the Planning Budgeting Committee and the Israel Science Foundation (grant NO 1937/12).

\pagebreak

\bibliography{Axino}

\end{document}